\documentclass[sigconf]{acmart}

\def\isarxiv{1}  

\newif\ifarxiv
\newif\ifnotarxiv
\ifdefined\isarxiv
\arxivtrue
\fi

\ifarxiv\notarxivfalse
\else\notarxivtrue
\fi

\usepackage{comment}
\usepackage{amsmath, amsfonts}
\usepackage{graphicx}
\usepackage{textcomp}
\usepackage{glossaries}
\usepackage{xcolor}
\def\BibTeX{{\rm B\kern-.05em{\sc i\kern-.025em b}\kern-.08em
    T\kern-.1667em\lower.7ex\hbox{E}\kern-.125emX}}
\usepackage{booktabs} 
\usepackage{makecell}
\usepackage{paralist}
\usepackage[inline]{enumitem}
\usepackage{svg}
\usepackage{placeins}
\usepackage{siunitx}
\usepackage{proof} 
\usepackage{bussproofs} 
\usepackage[frozencache,cachedir=_minted]{minted}
\usepackage{hyperref}
\usepackage{lipsum}
\usepackage{xpatch,letltxmacro}
\usepackage{syntax}
\usepackage{listings}
\usepackage{mathtools}
\usepackage{subcaption}
\usepackage{cleveref}
\usepackage{pifont}
\usepackage{caption}
\usepackage{xspace}
\usepackage{algpseudocode}
\usepackage{algorithm}
\usepackage{colortbl}
\usepackage{multirow}

\ifdefined\enableComments
  \newcommand{\authnote}[2]{({\bf \textcolor{blue}{#1}: \em \textcolor{red}{#2}})}
  \newcommand{\new}[1]{\textcolor{blue}{#1}}
\else
  \newcommand{\authnote}[2]{}
  \newcommand{\new}[1]{#1}
\fi

\newcommand{\gtan}[1]{\authnote{GT}{#1}}

\Crefname{section}{\S$\!$}{\S\S$\!$}

\newcommand{\code}[1]{\texttt{#1}}
\newcommand{\spec}[1]{\texttt{#1}}

\renewcommand{\paragraph}{\noindent\textbf}
\newcommand{\RNum}[1]{\uppercase\expandafter{\romannumeral #1\relax}}

\newcommand{\appname}[1]{\texttt{#1}}
\newcommand{\libname}[1]{\texttt{#1}}
\newcommand{\website}[1]{\textit{#1}}

\newcommand{\mimalloc}[0]{\libname{mimalloc}}
\newcommand{\toolname}[0]{\appname{SandCell}}
\newcommand{\sdrad}[0]{\appname{SDRaD}}
\newcommand{\newsdrad}[0]{\sdrad{}-v2}
\newcommand{\sandcrust}[0]{\appname{SandCrust}}
\newcommand{\sdradrustffi}[0]{\appname{SDRaD-FFI}}

\newcommand{\dCOne}{\ding{192}\xspace}
\newcommand{\dCTwo}{\ding{193}\xspace}
\newcommand{\dCThree}{\ding{194}\xspace}
\newcommand{\dCFour}{\ding{195}\xspace}
\newcommand{\dCFive}{\ding{196}\xspace}
\newcommand{\dCSix}{\ding{197}\xspace}
\newcommand{\dCSeven}{\ding{198}\xspace}
\newcommand{\dCEight}{\ding{199}\xspace}

\newcommand{\citeUnsafeIso}{~\cite{almohriFideliusCharmIsolating2018,riveraKeepingSafeRust2021,kirthPKRUsafeAutomaticallyLocking2022,liuSecuringUnsafeRust2020,bangTRustCompilationFramework2023}}
\newcommand{\cha}[1]{\hyperref[sec:motivation]{Challenge \RNum{#1}}}
\newcommand{\namedRef}[2]{\hyperref[#2]{#1~\ref*{#2}}}

\newacronym{ffi}{FFI}{foreign-function interface}
\newacronym{sfi}{SFI}{software-based fault isolation}
\newacronym{mpk}{MPK}{memory protection keys}
\usepackage{color}
\definecolor{eclipseBlue}{RGB}{42,0.0,255}
\definecolor{eclipseGreen}{RGB}{63,127,95}
\definecolor{eclipsePurple}{RGB}{127,0,85}

\lstdefinestyle{pagoda}{
  basicstyle=\fontfamily{FiraMono-TLF}\small,
  aboveskip={0.9\baselineskip},               
  keepspaces=true,
  numbers=left,
  escapechar=|,
}
\lstdefinelanguage{pagoda}
{
  morekeywords={sb_type,sb_enter,sb_spawn,sb_state,sb_impl,sb_struct},
  morekeywords={let,fn,struct,trait,impl,mut,if,else,return,for,while,loop,match,use,as,ref,move,enum},
  morecomment=[l]{//},
  morecomment=[s]{/*}{*/},
  morestring=[b]"
}

\begin{document}

\title{\toolname{}: Sandboxing Rust Beyond Unsafe Code}
\date{}

\ifarxiv
\settopmatter{printacmref=false} 
\renewcommand\footnotetextcopyrightpermission[1]{} 
\pagestyle{plain} 

\author{Jialun Zhang}
\affiliation{%
  \institution{Pennsylvania State University}
  \city{University Park}
  \country{USA}
}
\email{jialun.zhang@psu.edu}

\author{Merve Gülmez}
\affiliation{%
  \institution{Ericsson Security Research}
  \country{Sweden}
}
\email{merve.gulmez@ericsson.com}

\author{Thomas Nyman}
\affiliation{%
  \institution{Ericsson Product Security}
  \country{Sweden}
}
\email{thomas.nyman@ericsson.com}

\author{Gang Tan}
\affiliation{%
  \institution{Pennsylvania State University}
  \city{University Park}
  \country{USA}
}
\email{gtan@psu.edu}

\else
\renewcommand{\shortauthors}{Anon. Submission Id: 2275}
\global\def\authors{Anonymous Author(s)}
\acmBooktitle{first review cycle of CCS'26}
\acmConference[First review cycle of CCS'26]{}{November 15-19, 2026}{The Hague, The Netherlands}
\setcopyright{none}

\renewcommand{\shortauthors}{Zhang et al.}
\fi


\begin{abstract}
  Rust is a modern systems programming language that ensures memory safety by
  enforcing ownership and borrowing rules at compile time. While the unsafe
  keyword allows programmers to bypass these restrictions, it introduces
  significant risks. Various approaches for isolating unsafe code to protect
  safe Rust from vulnerabilities have been proposed, yet these methods provide
  only fixed isolation boundaries and do not accommodate expressive policies
  that require sandboxing both safe and unsafe code. This paper presents
  \toolname{} for flexible and lightweight isolation in Rust by leveraging
  existing syntactic boundaries. \toolname{} allows programmers to specify which
  components to sandbox with minimal annotation effort, enabling fine-grained
  control over isolation. The system also introduces novel techniques to
  minimize overhead when transferring data between sandboxes. Our evaluation
  demonstrates \toolname{}’s effectiveness in preventing vulnerabilities across
  various Rust applications while maintaining reasonable performance overheads.
\end{abstract}

\maketitle
\section{Introduction}\label{sec:intro}


Rust is a modern system programming language that provides memory-safety by
enforcing strict ownership and borrowing rules at compile time. However, in
practice, not all Rust programs fully benefit from compiler-enforced memory
safety. For example, pre-compiled Rust libraries or libraries written in other
programming languages must be called via \glspl{ffi} which the compiler cannot
analyze. Also, because the analysis is conservative, some valid,
memory-safe Rust programs will fail to type check. Rust provides an
\emph{unsafe} language superset that allows an ``escape hatch" from the
compiler-enforced safety analysis.

However, unsafe code is generally hard to get right and has become a common
source of bugs in Rust~\cite{rustfuzztrophy}. To write correct unsafe code,
programmers need to manually maintain the Rust ownership and borrowing rules and
other safety invariants expected by the Rust compiler in the unsafe code block.
For example, Rust developers must correctly annotate the lifetime of objects
beyond the \gls{ffi} boundary, and ensure any necessary preconditions necessary for
memory-safe execution of C functions are met. Failure to meet these expectations
can lead to memory-safety
vulnerabilities~\cite{nistCVE202524898,rustsecRUSTSEC20230044Openssl}.

Even without unsafe code, safe Rust is not immune to memory-safety issues. For
example, a long-existing soundness bug~\cite{rust-implied-bounds-bug} in the
type checking of \appname{rustc} (the official compiler of Rust) allows programmers to
arbitrarily expand the lifetime of a reference in safe Rust, which can lead to
dangling references. Similarly, the Rust standard library is generally
considered part of the safe Rust ecosystem. But given the common use of unsafe
code in the standard library, it might contain memory
vulnerabilities as well~\cite{githubBTreeSetCauses,githubFaultPushing,githubSoundnessIssue}.


Given these challenges, there is interest in mitigating lingering memory-safety
issues by compartmentalizing Rust programs. A natural idea is to isolate unsafe
code from safe code to prevent memory vulnerabilities from affecting the safe
subset of the program\citeUnsafeIso{}. These approaches categorize memory
objects into safe and unsafe categories according to how they are accessed.
Unsafe objects are then placed in a separate compartment, preventing bugs in
unsafe code from directly affecting safe objects.

These approaches are, however, limited in several ways. First, their isolation
boundary is fixed between safe and unsafe code. In some scenarios discussed in
\Cref{sec:motivation}, this boundary is not expressive enough to accommodate
fit-for-purpose isolation policies. Second, these approaches do not support
separate sandbox instances for different executions of the same piece of code;
all executions of the code would execute within one sandbox and share its
privileges and runtime data. As a result, they would not support, for example,
isolation between handling different requests in a web server, where a separate
sandbox is needed for each request, so one request's handler cannot interfere
with the handlers of other requests.

Third, by putting all safe code into the trusted computing base (TCB), all prior
approaches are inherently vulnerable to soundness bugs in the \appname{rustc} as
mentioned earlier. Furthermore, prior approaches consider the standard library
as part of the TCB, despite the presence of unsafe code in it.

\paragraph{This paper and contributions.}
To address these shortcomings of state-of-the-art compartmentalization solutions
for Rust, we propose \toolname{}, a compiler-based approach with three novel
properties. Firstly, to allow flexible isolation boundaries while minimizing
developers' effort on annotation, \toolname{} leverages syntactic boundaries
that already exist in the program (e.g., functions  and libraries, called
  ``crates" in Rust \footnote{Rust code is distributed as crates, similar to
packages or libraries in other languages.}) to let developers specify functions,
types,  or crates as sandboxes. We refer to sandboxes derived from such
syntactic boundaries as \emph{syntactic sandbox instances}. \toolname{}
automatically partitions a Rust application into distinct compartments according
to such syntactic instances and inserts sandbox-switching code to enforce the
isolation.

To address the second shortcoming, \toolname{} allows sandboxes to be
\emph{transient}; i.e., every time a certain piece of sandboxed code is invoked,
a new \emph{run-time instance} of the sandbox is created, so that different
execution instances of the same code are prevented from accessing each other's data.  This allows \toolname{} to isolate, for example, the handling of one web request from others requests in a web server.

Finally, \toolname{} creates isolation boundaries according to developer
specifications and does not distinguish between safe and unsafe code. This
ensures that the isolation policy defined using \toolname{} can effectively
confine exploits arising from soundness bugs in \appname{rustc} or memory
vulnerabilities in the standard library implementation.


In summary, we make the following contributions:
\begin{itemize}
  \item
    A compiler-based analysis allowing for users to specify syntactic instances of
    sandboxes for fine-grained and automated sandboxing of Rust functions, types,
    modules, and entire crates. We discuss the motivation and challenges in \Cref{sec:motivation} and the design in \Cref{sec:design}.

  \item
    Run-time instance sandboxing, which isolates different executions of the same
    code from each other. We motivate the need in \Cref{sec:motivation} and
    describe the design in \Cref{sec:design}.

  \item
    \toolname{}, a general-purpose and highly-automated sandboxing tool for Rust
    supporting both syntactic instance and run-time instance sandboxing.

  \item We evaluate \toolname{} on several real-world Rust
    applications, demonstrating its ability to mitigate vulnerabilities and
    expressing various isolation patterns with reasonable overheads in \Cref{sec:eval}.
\end{itemize}


\section{Background}\label{sec:background}

This section gives background on Rust, including its memory safety guarantees
and memory-unsafe language features; and also discusses the concept of
in-process isolation, which is used as our low-level isolation mechanism.

\subsection{Memory Safety in Rust}

Safe Rust guarantees memory safety by enforcing strict ownership and borrowing
rules at compile time. Informally, it has three rules:
\begin{enumerate*}
\item Every value has a single owner; assignment to a new owner invalidates the
  previous owner.
\item At any time, a value can be either modified through one reference (either
  by the owner or through a single mutable reference) or simultaneously read
  through multiple immutable references, but not both.
\item No reference can outlive the value it points to.
\end{enumerate*}
To enforce the third rule, Rust associates a \emph{lifetime} with each reference
at the type-system level, which can be viewed as a set of program points where
the reference is valid. The Rust compiler can infer the lifetimes of most
references, and the rest must be explicitly annotated by the programmer.

The Rust compiler statically enforces these rules to prevent common memory
vulnerabilities such as use-after-free and double-free bugs. However, due to the
conservative nature of Rust's static analysis, some valid (i.e., memory-safe)
programs might be rejected by Rust's borrow checker that
enforces ownership and borrowing rules at compile time.

\subsection{Unsafe Rust}

In Rust, the \emph{unsafe} keyword serves two high-level purposes. First, annotating a
function definition with unsafe indicates that the function has contracts that
the Rust compiler cannot verify. Second, annotating a function call or code
block with unsafe indicates that the programmer has manually ensured the
contract is satisfied. In practice, programmers use unsafe code for different
reasons, and here we discuss the two most common
ones~\cite{astrauskasHowProgrammersUse2020}.

Around 90\% of the unsafe code in Rust is used to call unsafe
functions~\cite{astrauskasHowProgrammersUse2020}. For example, calling a version of array indexing that does not perform bounds checking is unsafe, since the programmer is responsible for ensuring that the index is within
bounds. Another example is calling a function from an external library written
in C. In these cases, the developer must ensure that preconditions expected by
the unsafe function are satisfied. Violating these preconditions can lead to
bugs such as RUSTSEC-2023-0044~\cite{rustsecRUSTSEC20230044Openssl}, which is a
vulnerability from \libname{rust-openssl} where a string from Rust code is
passed to C code without null-termination, causing an out-of-bound read.

Another common use case of unsafe code is to manipulate raw pointers. It might
cause bugs when, for example, the developer assigns an incorrect lifetime to a raw
pointer that is cast to a reference. Since the lifetime of a raw pointer is not
tracked by the compiler, such bugs cannot be detected at compile time. As a
practical example, CVE-2025-24898~\cite{nistCVE202524898} is a vulnerability in
\libname{rust-openssl} where an incorrect lifetime parameter is bound to a buffer
(allocated in the C code). Because the lifetime parameter exceeds the actual
lifetime of the buffer, it leads to a use-after-free vulnerability.

\subsection{Bugs in Safe Rust}\label{sec:bugs-in-safe-rust}

The official Rust compiler, \appname{rustc}, is not bug-free either and might
accept programs that violate its safety invariants. For example, a well-known
bug~\cite{rust25860} in the type checker enables the lifetime
of a local reference to be expanded to the global lifetime. Exploiting such soundness bugs, adversaries may intentionally upload vulnerable crates to \website{crates.io}, the Rust package index, to
mount supply-chain attacks. One tool Rust developers have at their disposal to
minimize the risk posed by third-party dependencies is limiting the dependencies
to crates without unsafe code. The package index tags such crates with the
``no\_unsafe" marker to make finding such crates easier. This mechanism
incentivizes malicious developers to exploit soundness bugs instead of unsafe
code. The \appname{rustc} team maintains a list of such compiler
bugs~\cite{rust-unsound-bug-list}.

The Rust standard library is also an important building block for safe Rust code
as it provides safe abstractions for low-level operations such as heap
allocation and synchronization primitives. But bugs in the standard library also
exist. For example, an integer overflow led to a buffer overflow in
\code{std::str::repeat}~\cite{RustStdCVE20181000810}.
\subsection{In-process isolation}
One way to sandbox cooperating software modules is to place
each in a process with its own virtual address
space~\cite{lamowskiSandcrustAutomaticSandboxing2017}. However, for
tightly-coupled modules, process-based isolation incurs prohibitive context
switch and communication overhead~\cite{gulmezFriendFoeExploring2023}.
\emph{In-process isolation} is a security technique that places the code and
data for a sandboxed module in its own \emph{fault domain}, a logically separate
portion of the application's address space.  In-process isolation trades strong,
hardware- and OS-enforced process isolation for substantially faster context
switches between fault domains. In-process isolation can be realized portably
and language-independently through \gls{sfi}~\cite{wahbe1993efficient,Tan17}. \gls{sfi}, however, increases
execution times for the sandboxed modules as their memory accesses must be
checked for validity by a software-enforced \emph{in-line reference monitor}.
Modern approaches for in-process isolation leverage hardware features, such as
\gls{mpk}, to improve upon the efficiency of \gls{sfi}.  The
popularity of sandboxed JavaScript in modern browsers, managed languages, and
application sandboxing in mobile OSes have motivated processor manufacturers to
include such hardware features in commercial, off-the-shelf processors.

\hyphenation{hardware-assisted}
\paragraph{Protection Keys for Userspace.}
\gls{mpk} on the x86 architecture uses the \emph{protection keys for
userspace} (PKU) hardware feature.  PKU associates every 4KB memory page with a
4-bit protection key identifier which is stored in the OS's page table entry.  A
32-bit CPU register, the \emph{protection key rights register} (PKRU), controls
the access rights of a  particular key.  The PKRU register holds a 2-bit mask
for each of the 16 protection keys.  The state of the bits specifies whether
pages corresponding to the key are writable/accessible.

In contrast to OS-enforced paging, the PKRU configuration can be modified by the
userspace process.  Since PKU hardware enforces access control based on the
current configuration of the PKRU, unauthorized modification of the PKRU, or
techniques that bypass PKU enforcement completely makes PKU vulnerable to
threats from run-time attacks and malicious software aiming to escape from its
sandbox.  For example, an attacker could manipulate \code{/proc/self/mem} to
bypass \gls{mpk} protections~\cite{connor20} or misuse the \code{wrpkru} instruction
to modify the PKRU register~\cite{Vahldiek-Oberwagner19}.

\begin{figure}[t]
\begin{minipage}{.40\textwidth}
\centering

\begin{minted}[escapeinside=||, xleftmargin=1pt]{rust}
// Function from C library
extern "C" { 
  fn parse(req: *const u8) -> *const u8;
}

use crypto;

mod request {
  fn handle_request(
    &self, req: &Socket
  ) -> Result<(), Error> {
    let data = req.get_data()?;
    let data = unsafe { parse(req) }; |\label{bg:unsafecall}|
    let key = crypto::get_key()?;
    crypto::verify(&data, &key)?;
    let res = send(&data, req);  
    Ok(())
  }
}

fn main() {
  let socket = /* omitted */;
  loop {
    let req = accept(socket);
    match handle_request(req) {
      Ok(()) => continue,
      Err(e) => handle_error(e),
    }
  }
}

\end{minted}
\end{minipage}
\caption{A simplified web server. In Rust, modules (the keyword \code{mod}) are used to organize definitions such as functions and types.} \label{fig:mot}

\end{figure}

\section{Motivation and Challenges}\label{sec:motivation}

Given the disposition of vulnerabilities in unsafe Rust code to undermine the
compiler-enforced memory-safety properties, a natural idea is to isolate
unsafe code from the rest of a Rust program, as explored by previous
work\citeUnsafeIso{}. However, simply having two compartments (one safe and one unsafe compartment) is insufficient for the reasons discussed in \Cref{sec:intro} and \Cref{sec:background}. Consequently, compartmentalization in Rust must solve the following
challenges.



\paragraph{Challenge \RNum{1}: Isolation between unsafe
code.} Consider a web server sketched in \autoref{fig:mot}
that handles requests by first parsing the request and then verifying it using a
cryptographic library. The parsing and cryptographic libraries are both written
in C. In particular, the \code{parse} function is directly called via \glspl{ffi}, while the cryptographic library is wrapped inside a
Rust crate \code{crypto}.

If we simply put all unsafe code into a separate compartment, a potential memory vulnerability
inside the \code{parse} function can still be exploited to access the memory of
the \code{crypto} library, which also contains unsafe code. Therefore, what is more desirable is to put these two libraries into two different sandboxes. This
requires the ability to specify isolation boundaries at the library granularity,
while existing approaches support only a predefined isolation boundary
between all safe and all unsafe code. More generally, the isolation tool should allow
developers to isolate different syntactic instances such as functions, modules,
and crates.

\paragraph{Challenge \RNum{2}: Isolation between different executions of the same
code.} Consider another attack scenario where the
adversary sends a malicious request to exploit a use-after-free bug in the
\code{crypto} library to belonging to other requests that remains in memory. To counter this, a policy isolates the code handling distinct requests from each other is needed.
This, however, cannot be accommodated by approaches that sandbox only unsafe
code, which shares the states of all requests in a single unsafe-code sandbox. In a broader
sense, \cha{2} requires the ability to allow the developer to isolate instances of the same code that execute concurrently at run-time from each other.


\paragraph{Challenge \RNum{3}: Isolating soundness bugs and bugs in standard
library.} The Rust compiler itself is not bug free and
might accept programs that violate its safety invariants, as discussed in
\Cref{sec:bugs-in-safe-rust}. By sandboxing untrusted Rust code, the impact of
such bugs can be limited to the sandboxed code and data. Such cases requires the ability to isolate a mix of safe and unsafe code.

Rust's standard library is another important building block for its safety
guarantees and provides safe abstractions for low-level operations such as
heap allocation and synchronization primitives. Previous approaches that isolate
unsafe code from safe code generally include the standard library in the TCB.
To mitigate the impact of bugs in the Rust standard library (e.g., \cite{RustStdCVE20181000810}), the
compartmentalization method must support sandboxing not only untrusted and
user-specified code, but also dependencies and their transitive dependencies.

\paragraph{Challenge \RNum{4}: Reducing data exchange cost.}\label{challenge:datacopy}
Frequent data exchanges between sandboxes can become a performance bottleneck. The common approach of deep copying data between sandboxes is especially costly as it needs to copy all transitively reachable data.



\section{System and Adversary Model}\label{sec:threat}


Our system model is similar to other in-process compartmentalization approaches:
we assume that the protected application starts in a trusted state from which new,
untrusted compartments of sandboxed code can be created at runtime. Similarly to
prior work, we also assume that the compartmentalization policy is decided at compile
time and remains immutable throughout the lifetime of the application.

We assume an adversary who exploits either pre-existing vulnerabilities in Rust
applications, or intentionally introduces vulnerabilities they can exploit into
crates that they control. Using such vulnerabilities, the adversary can mount
control-flow or data-only attacks against the victim application. The adversary
may also attempt to misuse privileges granted to the application, such as
performing unwanted, harmful system calls. The goal of \toolname{} is to
mitigate such attacks by isolating untrusted code and data from the rest of the
program.
However, we do
not provide security guarantees against attackers specifically targeting the user-space system-call interpositioning (see ~\Cref{sec:syscall-filter}).
It is also possible that the attacker can indirectly
impact the program through information flow between trusted and
untrusted code such as through side channels, which is not in the scope of \toolname{}.


\section{Design}
\label{sec:design}

In this section, we give an overview of the design of \toolname{}, including the
overall workflow, user specification, its semantics, and the run-time
architecture. We will delay the discussion of the program analysis supporting
\toolname{} until \Cref{sec:dataflow-analysis} and the implementation details
until \Cref{sec:impl}.

\subsection{Overview}
\label{sec:overview}

\toolname{} takes the Rust source code and a sandboxing specification as input,
and produces a sandboxed binary as output. The \toolname{} specification
contains a list of functions, types, 
or crates to be sandboxed. 
We will elaborate on this in~\Cref{sec:user-spec}. \toolname{}'s analysis
builds a call graph of the program and uses the specification to identify the
instrumentation points necessary for sandboxing the specified components.
\toolname{} also uses flow-insensitive data-flow analysis to identify allocation
sites that might allocate heap objects that are accessed across sandbox
boundaries (\Cref{sec:dataflow-analysis}). These results are used by the
instrumentation component (\Cref{sec:instrumentation}) of \toolname{} to alter
the program's source code and insert calls into the low-level isolation
application programming interfaces (API) at the instrumentation points
identified during the analysis. The in-process isolation APIs and the run-time
architecture of \toolname{} will be described in \Cref{sec:runtime-arch}.
Finally, the instrumented program is compiled by \appname{rustc} as normal,
producing a sandboxed binary.

\subsection{User Specification}
\label{sec:user-spec}

\begin{figure}[t]
\centering
\begin{minted}[linenos=false]{c}
[functions] // can be [types] or [crates]
foo = { transient = true }
bar = { transient = false }
\end{minted}
\caption{A \toolname{} specification example.} \label{fig:spec}
\end{figure}

In this section, we describe the syntax and semantics of the \toolname{}
specification and how they address the challenges discussed in
\Cref{sec:motivation}.

To address \cha{1}, \toolname{} allows users to define sandboxes for functions,
modules, crates, and types by listing their names in the specification file, as
illustrated by the example in \Cref{fig:spec}, which specifies that
functions \code{foo} and \code{bar} should be sandboxed. When there are multiple
candidates with the same name, users can refine it with quantified paths.

In response to \cha{2}, we associate a \spec{transient} field with every
sandbox. If \spec{transient} is set to \spec{true}, every call to the function
will enter a freshly created sandbox. Otherwise, all calls to the function will
share the same sandbox instance; i.e., the code and data of the function will be
shared across different calls.

For \cha{3}, \toolname{} does not distinguish between safe and unsafe code, but
instead follows the user-specified sandboxing boundaries. Therefore,
vulnerabilities in the Rust standard library or soundness bugs in
\appname{rustc} can be mitigated by sandboxing the code that uses them.

For \cha{4}, to avoid the high cost of deep copying data between sandboxes, we employ a novel approach that utilizes static analysis to compute allocation sites that may allocate cross-sandbox data, and then instruments those allocation sites so that the data is allocated in some shared data domain between sandboxes that exchange data.


Invocations of different units (functions, crates, etc.) create different
sandboxes. For the example in \autoref{fig:spec}, invocations of functions
\code{foo} and \code{bar} create different sandboxes; so their executions are
isolated from each other. Conceptually, a sandbox is a memory region
that contains the code and data of the sandboxed unit (function, crate, etc.) as
well as its transitive dependencies. For example, if both \code{foo} and
\code{bar} calls another function \code{baz}, there is a separate copy of the
code and data of \code{baz} in each sandbox.

For modules and crates, all publicly accessible functions in the module/crate
will be sandboxed in the same way as above, while sharing the same sandbox
domain. Note that nested calls to the same module/crate will not enter a new
sandbox domain. For example, for a sandboxed crate \code{C} with a public
function \code{C::foo} and a private function \code{C::bar} that is called by
\code{C::foo}, calling \code{C::foo} from the trusted domain will enter the
sandbox domain of \code{C}, while the call to \code{C::bar} within \code{C::foo}
stays in the same sandbox domain as a normal function call. For sandboxed
types, all associated methods of the type that are publicly accessible will
share the same domain. However, if the \spec{transient} field is set to
\spec{true}, a fresh sandbox domain will be used for each call to a publicly
accessible function of the type/
crate.

We assume that different instances in the specification do not overlap with each
other in the syntax of the program. For example, if a function \code{foo} and a
crate \code{c} are both specified in the specification, the function \code{foo}
should not be a member of \code{c}. \toolname{} supports nested sandbox calls so
that the code in a sandbox can invoke functions in another sandbox. \toolname{}
could support nested sandbox creation, but we did not find any compelling use
case. Thus, our implementation allows only the main sandbox to create new
sandboxes.

\new{\paragraph{Specification strategies.}}
\new{In \toolname{}, programmers specify sandbox boundaries using syntactic
  boundaries of program units such as functions and crates. Identifying good
  isolation boundaries is a well-studied problem: tools such as
  Privtrans~\cite{brumley2004privtrans}, SeCage~\cite{liu2015thwarting},
  PtrSplit~\cite{liu2017ptrsplit}, and Program-mandering~\cite{programMandering}
  can automatically compute sandbox boundaries based on information flow analysis
  or sensitive data annotations. However, these automatic approaches typically
  enforce fixed kinds of policies (e.g., protecting confidential information). On the
other hand, developers with deep knowledge of a codebase understand the program's behavior and trust relationship and can thus develop  policies targeting custom security properties.}

\new{Since \toolname{} is orthogonal to both approaches, it can be combined with
  automatic partitioning tools or expert-crafted policies. However, even without
  such tools or expertise, developers can follow simple strategies that provide
meaningful security benefits:}
\begin{itemize}[leftmargin=*]
  \item \textbf{Library-based (supply-chain):} Sandbox entire third-party
    libraries, especially those with known vulnerabilities or written in unsafe
    languages. This strategy mitigates supply-chain attacks without requiring
    understanding of the library's internals.

  \item \textbf{User-based:} In multi-tenant applications, isolate components
    that process different users' data from each other. This prevents a
    compromised user session from attacking other users.

  \item \textbf{Input-oriented:} Sandbox components that directly handle
    external input, such as parsers, decoders, and deserializers. These components
    are common attack surfaces and benefit from isolation regardless of whether
    specific vulnerabilities are known.
\end{itemize}

\new{In our case studies, we found that trust boundaries align naturally with
  syntactic boundaries (crates, modules, functions), making specification
straightforward. We evaluate these strategies in \Cref{sec:eval}.}
\subsection{Analysis and Instrumentation}\label{sec:analysis-and-inst}

After the user specifies the sandbox boundaries, the next step is to insert
necessary code for enforcing sandboxes and passing data between sandboxes. Given
the source code, \toolname{} builds a call graph and a data flow graph of the
program. Based on the specification and the call graph, \toolname{} identifies
the public interfaces of types, modules and crates to be sandboxed. Then it
inserts calls to the low-level isolation APIs at these interfaces to enforce the
isolation at runtime.

For low-level isolation, \toolname{} adopts an open-sourced isolation library
\sdrad{} by Gülmez et al~\cite{gulmez2023dsn}, and augments it as \newsdrad{},
adding syscall filtering based on \libname{zpoline}~\cite{yasukata23} and
\mimalloc{}~\cite{Leijen2019MimallocFL} as the memory allocator. More details
can be found in \Cref{sec:in-process}. At a high level, \newsdrad{} provides a
set of APIs for managing sandboxes. To switch between sandboxes, \toolname{}
inserts calls to these APIs at the entry and exit points of the sandboxed
functions. Also, it provides APIs for management of shared data domains for
allocation of objects that two sandboxes can access. More details on the
\newsdrad{} APIs can be found in \Cref{sec:impl}.

The data passed across sandbox boundaries includes the arguments passed to the
sandboxed functions and their return values. To make data from one sandbox
available to another sandbox, one approach is to deep copy all such data as in
Glamdring~\cite{lind2017glamdring}, PtrSplit~\cite{liu2017ptrsplit}, and
GOTEE~\cite{ghosn2019secured}. This approach, however, can lead to a high
overhead since deep copying copies all transitively reachable data; it should be
avoided when the size of transmitted data is
large~\cite{gulmezFriendFoeExploring2023}.

Instead of deep copying, our design copies only stack data and allocate heap
data (when it does not contain references to stack data) on a shared data domain
that is accessible from both sides of a cross-boundary call. This approach
reduces the overhead of copying cross-boundary data. On the other hand, it also requires that cross-boundary data may not be allocated next to each other
the relevant memory pages can be assigned to a shared memory region. If we
have to copy such data to a shared data domain during cross-boundary calls, we
are back to the same problem of high copying cost. To address this, we propose a
novel approach to
\begin{enumerate*}[label=(\arabic*)]
\item employ static analysis to compute allocation sites that
  may allocate cross-boundary data in memory, and
\item instrument those allocation sites so that the data is allocated in a
  shared data domain. This approach avoids the data copying cost by directly
  allocating cross-boundary data in a shared domain.
\end{enumerate*} We
detail the static analysis in \Cref{sec:dataflow-analysis} and the
instrumentation in \Cref{sec:impl}.

\subsection{Run-time Architecture}\label{sec:runtime-arch}

\begin{figure}[t!]
  \centering
  \includegraphics[width=\linewidth]{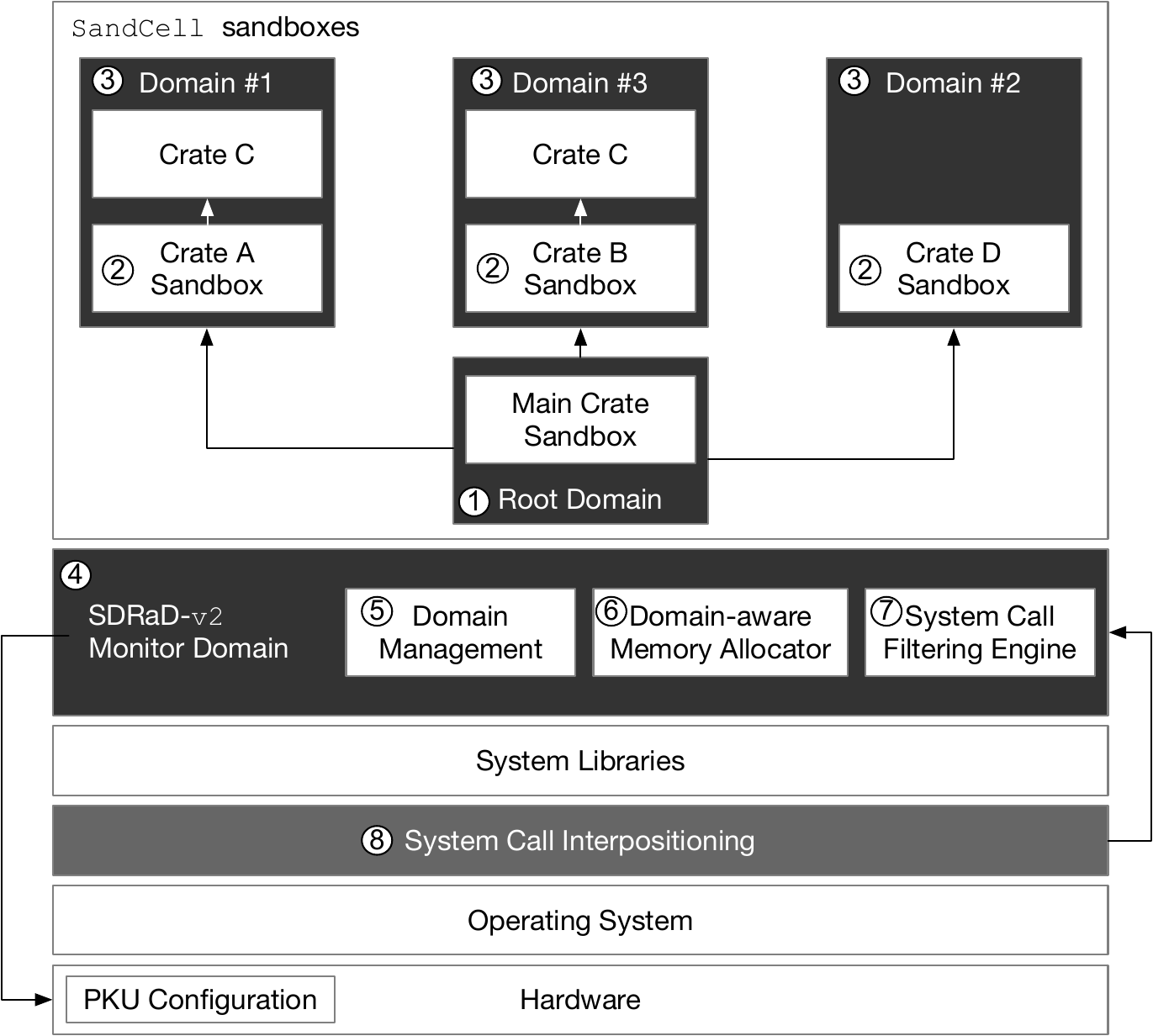}
  \caption{\toolname{} run-time architecture.}\label{fig:runtime-arch}
\end{figure}

\Cref{fig:runtime-arch} shows an overview of the \toolname{} architecture at
run time. When the program starts, \newsdrad{} places the main function and the
program stack and heap into a special \emph{root domain} (\dCOne).  When
\toolname{} creates a sandbox (\dCTwo), a new \newsdrad{} domain (\dCThree) is
initialized and a new, dedicated, stack and heap area is allocated.  Code in the
\newsdrad{} root domain can access data in any new domain created by \toolname{},
but sandboxed code cannot access the stack or heap of other domains, including
the root domain.  \newsdrad{} control data is stored with a separate, \emph{monitor
domain} (\dCFour) that is inaccessible to code in the root or sandboxed domains.
The monitor domain will run the code for domain management (\dCFive), the
domain-aware memory allocator (\dCSix), and the system call filtering
(\dCSeven). We will discuss the low-level details in \Cref{sec:impl}.

Notice that, as shown in \Cref{fig:runtime-arch}, crate \code{C} is a dependency
of both crate \code{A} and crate \code{B}. However, at run time, because crate
\code{A} and crate \code{B} are sandboxed separately, each of them has its own
copy of the code and data of crate \code{C}.

\subsection{Security Guarantees}

{\toolname{} provides memory isolation between sandboxes, preventing memory
  vulnerabilities from being exploited to access memory outside the sandbox. Suppose that in \Cref{fig:runtime-arch}, the attacker exploits a
  memory vulnerability in crate \code{A} and gains the ability to read and write
  arbitrary memory. Because of the memory isolation provided by \toolname{}, the
  attacker can only directly manipulate the memory within Domain \#1. Attempts to
  access memory in other domains, including the root domain, will result in a
  runtime error. As a practical example, CVE-2018-25032~\cite{CVE-2018-25032} is a out-of-bound
  write triggered by crafted user input in \libname{zlib}. If sandboxed by
  \toolname{}, the out-of-bound write can only modify memory within the sandbox.
  Another example, CVE-2023-0216~\cite{CVE-2023-0216}, is an invalid pointer
  dereferencing vulnerability in \libname{openssl}. Sandboxing the
  library confines the invalid memory access within the sandbox.
}

We note that the guarantee provided by \toolname{} is typical for sandboxing tools and it does not provide \emph{interface safety}~\cite{SandboxingSok2025SP}, which requires the data and control flow at the sandbox boundaries be carefully validated
to prevent potential misuses or abuse. How to provide interface safety effectively and efficiently in sandboxing is still an area of research.


\section{\new{Dataflow Analysis}}
\label{sec:dataflow-analysis}

We next describe the static analysis that identifies which heap objects
may cross sandbox boundaries. We first illustrate the problem with a motivating
example, then describe our analysis algorithm.

\subsection{Problem definition}

\paragraph{Goal.}
When a sandboxed function is called, any heap data passed as arguments must
reside in a memory region accessible to both the sandbox and the main program.
Our analysis identifies these \emph{cross-boundary heap objects} at compile
time, so their allocation sites can be instrumented to use the shared memory
domain.

\paragraph{Motivating example.}
Consider a common pattern where data is accumulated through helper functions, illustrated by the following example, where the main function first loads a series of plugins and then invokes each plugin through a sandboxed call.
\begin{lstlisting}[numbers=none]
fn load_plugins() -> Vec<Plugin> {
    let mut plugins = Vec::new(); // Allocation
    for path in discover_paths() {
        if let Some(p) = parse_plugin(path) {
            plugins.push(p); // Container insert
        }
    }
    plugins
}
fn run_sandbox(plugins: &[Plugin]) {
    for p in plugins.iter() {
        sandbox(p); // Sandboxed call
    }
}
fn main() {
    let plugins = load_plugins();
    run_sandbox(&plugins);
}
\end{lstlisting}

The analysis must determine that allocation in \code{load\_plugins} flows
into the sandboxed call in \code{run\_sandbox}. This requires tracking data
through: (1)~a return value flowing from \code{load\_plugins} to \code{main};
(2)~a function argument flowing from \code{main} into \code{run\_sandbox};
(3)~an iterator \code{plugins.iter()} that borrows the container; and
(4)~the \code{push} operation that inserts elements into the vector.

Without interprocedural analysis, we cannot connect allocations in
\code{load\_plugins} to uses in \code{run\_sandbox}. Without modeling container
operations, we cannot determine that data pushed into \code{plugins} is
accessible through iteration. Our analysis handles these cases by building a
dataflow graph that is interprocedural and includes summary edges for
standard library container methods.

\paragraph{Identifying heap objects.}
In Rust, heap allocations are wrapped in safe abstractions like \code{Vec},
\code{HashMap}, \code{Box}, etc. Our analysis recognizes allocations by
detecting constructor calls for these standard library containers (e.g.,
\code{Vec::new()}, \code{HashMap::with\_capacity()}).

\subsection{Analysis algorithm}

\begin{figure}[t]
\centering
\small
\begin{tabular}{@{}l@{\quad}l@{}}
\toprule
\textbf{Node Type} & \textbf{Description} \\
\midrule
Local variable & A variable declared in a function \\
Parameter & A function's input parameter \\
Return value & A function's return value \\
Field projection & A struct field access (e.g., \code{x.field}) \\
Dereference projection & A pointer dereference (e.g., \code{*ptr}) \\
Heap allocation & A call to \code{Vec::new()}, \code{Box::new()}, etc. \\
\midrule
\textbf{Edge Type} & \textbf{Created For} \\
\midrule
Assign & Direct assignment: \code{x = y} \\
Ref & Reference creation: \code{x = \&y} \\
Aggregate & Struct/tuple construction \\
CallArg & Argument $\to$ parameter at call sites \\
Return & Return value $\to$ destination at call sites \\
StdSummary & Standard library function summaries \\
\bottomrule
\end{tabular}
\caption{Node and edge types in the dataflow graph. 
} \label{fig:mir-syntax}
\end{figure}

Our analysis operates on Rust's Mid-level Intermediate Representation (MIR),
which the compiler generates before optimization. We construct an
interprocedural \emph{dataflow graph} (DFG) that captures how values flow
through the program, then perform backward reachability to find allocation sites.

\paragraph{Dataflow graph structure.}
The DFG is a directed graph where:
\begin{itemize}
  \item \textbf{Nodes} represent program locations that hold values: local
    variables, function parameters, return values, struct fields, and heap
    allocation sites.
  \item \textbf{Edges} represent data flow from sources to destinations. An edge
    from node $A$ to $B$ means the value at $A$ may flow to $B$.
\end{itemize}
\Cref{fig:mir-syntax} summarizes the node and edge types in our DFG.

We create edges for the following program constructs:

\begin{enumerate}
  \item \textbf{Assignments} (\code{x = y}): Add an \emph{Assign} edge from \code{y} to \code{x}.
  \item \textbf{References} (\code{x = \&y}): Add a \emph{Ref} edge from \code{y} to \code{x}. \gtan{It's not from x to y?}
    We also handle reborrowing patterns like \code{x = \&*y}, where we first
    create a projection node for \code{*y}, then add a \emph{Ref} edge to \code{x}.
  \item \textbf{Dereferences} (\code{x = *y}): Add an \emph{Assign} edge from the
    projection node \code{*y} to \code{x}. The dereference node is connected to
    the pointer's base via an \emph{Aggregate} edge.
  \item \textbf{Struct/tuple construction} (\code{s = S \{ f: v \}}): Add an
    \emph{Aggregate} edge from \code{v} to the field node \code{s.f}, and another
    \emph{Aggregate} edge from \code{s.f} to \code{s}.
  \item \textbf{Function calls} (\code{r = f(a, b)}): Add \emph{CallArg} edges from
    each argument to the corresponding parameter of \code{f}, and a \emph{Return}
    edge from \code{f}'s return value to \code{r}.
  \item \textbf{Heap allocations} (\code{v = Vec::new()}): Create a heap allocation
    node and add an \emph{Assign} edge to \code{v}.
\end{enumerate}

For standard library functions whose bodies are not available or too complex to
analyze, we use pre-defined \emph{summaries} to specify which arguments flow
to the return value. For example, \code{Vec::iter()} returns an iterator over
its receiver; so we add a \emph{StdSummary} edge from the receiver to the result.

\paragraph{Handling indirect calls.}
Function pointers and trait objects create \emph{indirect calls} where the
target is not statically known. The analysis tracks how function pointers flow
through the DFG and uses this information to resolve possible call targets.
Critically, DFG construction and call target resolution are performed
\emph{iteratively}: we first build an initial DFG using only direct calls, then
resolve indirect call targets based on dataflow, add edges for newly discovered
calls, and repeat until no new targets are found. This iterative refinement
ensures that the call graph and DFG are mutually consistent.

\paragraph{Backward reachability.}
Given the DFG, we find allocation sites by traversing edges \emph{backward}
(from destination to source) starting from sandboxed call arguments.
The algorithm is a straightforward worklist-based graph traversal:
we initialize the worklist with all arguments of sandboxed function calls,
then repeatedly pop a node, follow its incoming edges, and add unvisited
source nodes back to the worklist.
When a heap allocation node is reached, we record it as a cross-boundary allocation.
The traversal continues until the worklist is empty.

\paragraph{Context sensitivity.}
The backward traversal is \emph{call-site sensitive}: when traversing a
\emph{Return} edge from callee to caller, we record the call site on a stack.
Later, when traversing a \emph{CallArg} edge, we check that it matches the
recorded call site. This prevents spurious paths that would ``return'' to a
different caller than the one that made the call. The analysis uses a
configurable stack depth limit to bound analysis time while maintaining
precision for common patterns.

\section{Implementation}
\label{sec:impl}

In this section, we provide the implementation details deferred in the previous
sections.

\subsection{Program Analysis}\label{sec:analysis-impl}

\toolname{}'s analysis is implemented as a plugin to \appname{rustc}. The
compiler translates Rust source code through several levels of intermediate
representations (IRs). The Mid-level IR (MIR) is the last IR before translation
into the language-neutral LLVM IR. MIR has a much simpler syntax than the
surface syntax of Rust but still retains variable type information, which will
be stripped away during ``\textit{lowering}'', i.e., translation to LLVM IR. All
\toolname{} analysis are performed at the MIR level.

In MIR, each function has a list of \code{Local}s, including the function
arguments, temporaries, local variables, and the return value. Each \code{Local}
might contain a list of \code{Place}s that correspond to memory locations, e.g.,
fields of a struct. We map each \code{Place} to a node in the dataflow graph and
create edges between \code{Place}s on the left-hand side and right-hand side of
an assignment as discussed in~\Cref{sec:dataflow-analysis}. The rest of analysis
has been described in \Cref{sec:dataflow-analysis}.

\subsection{Instrumentation}\label{sec:instrumentation}

The instrumentation phase inserts code to enforce the sandbox boundaries
specified by the user. For each function that crosses the sandbox boundary, a
wrapper function is inserted to switch the context to the callee's sandbox by
calling \newsdrad{} APIs, which will be elaborated in \Cref{sec:in-process}.
Then the instrumentor redirects the original function call to the wrapper
function.

For sharing heap objects, the instrumentor modifies the allocator of the heap
object using the standard library's allocator API. In Rust's standard library,
heap objects are generic over allocators; e.g., \code{Vec<T, A: Allocator>} can
be  instantiated with different allocators. By using \code{Vec<i32,
SandCellAlloc>}, we direct all memory allocations associated with \code{Vec} to
our custom allocator \code{SandCellAlloc}, which allocates memory in a shared
data domain. For types that do not support custom allocators, \toolname{} wraps
their allocations and all the following mutations with a macro that redirects
the allocations to the shared data domain.

After the analysis, we statically know how many sandboxes we need in an
application and what objects will be shared between which sandboxes. Then for
each combination of sandboxes that share data, a shared data domain is allocated
and a static ID is assigned to that domain. For example, if the analysis
determines that sandbox A, B, and C share data, then a shared data domain is
assigned to that combination.\footnote{It is worth noting that in our use cases
all shared data are between two sandboxes.} At allocation sites that allocate
objects for a shared data domain, the instrumentor passes the domain's static ID
to \toolname{}'s \code{SandCellAlloc} allocator so that those objects are
allocated in that particular data domain.

For stack objects, the instrumentor generates code to copy the stack object to
the callee's stack, following its type information. For references, the
instrumentor generates code to copy the referenced object.

\subsection{In-Process Isolation}\label{sec:in-process}

The low-level isolation mechanism used by \toolname{} is responsible for
enforcing the isolation between different sandboxes and providing primitives to
transition securely from one sandbox to another. \toolname{} implements
hardware-assisted in-process isolation via the x86-64 processor PKU feature by
building on an open-source isolation library, \emph{Secure Rewind \&{} Discard}~(\sdrad{})~\cite{gulmez2023dsn}, that
implements lightweight primitives for in-process isolation and fault recovery.
We augment \sdrad{} with support for a Rust-compatible memory allocator and
harden it by incorporating a system call filtering engine. For clarity, we
refer in the following description to \sdrad{} when functionality is provided by
the original \sdrad{} library, and to \newsdrad{} when describing added
functionality.

\paragraph{Secure Rewind and Discard}. Each \sdrad{} domain must be assigned a
unique ID when they are initialized using the \code{sdrad_init} API call.
\toolname{} is responsible for assigning an ID to each sandbox, and providing
the corresponding ID to the \code{sdrad_enter(id)} call that is used to enter an
initialized domain. Since \toolname{}'s isolation boundaries are all function
calls, \toolname{} inserts \code{sdrad_enter(id)} calls around sandboxed
functions at call sites where \sdrad{} needs to switch between caller's and
callee's domains. A corresponding \code{sdrad_exit()}
call marks the end of code execution, and is used to mark transitions back to
the caller's domain after a sandboxed function returns. \sdrad{} supports
multi-threading by intercepting calls to \code{pthread_create} and
\code{pthread_exit} to manage thread-specific isolation. During a domain
switch, \sdrad{} changes the stack pointer to point to the stack of the domain
that is being entered.  While executing inside a sandbox, allocations via the
POSIX \code{malloc(size)} API are redirected to use the heap that belongs to the
current \sdrad{} domain. Allocations that occur outside a domain, but should
reside on a specific domain's heap need to use the \code{sdrad_malloc(id, size)}
and \code{sdrad_free(ptr)} API that take a domain ID as an additional argument.
Any memory access to data that is not within a sandbox domain is blocked by
\sdrad{}'s isolation. \sdrad{} lacks PKU hardening features such as binary
rewriting of unauthorized \code{wrpkru} instructions or system-call filtering.

\paragraph{Changes in \newsdrad{}}. To ensure that heap-allocated objects are
placed into memory belonging to a particular domain, \sdrad{} provides a
domain-aware memory allocator built on the Two-Level Segregated Fit
(\libname{TLSF}) heap allocator~\cite{Masmano04}.  However, \libname{TLSF}
implementation used in \sdrad{} only supports 8-byte alignment for allocations,
whereas certain common Rust crates, such as \libname{hashbrown}, used by the
cargo test and cargo bench tools expect allocations to be made at 16-byte
alignment. \newsdrad{} replaces \libname{TLSF} with the \mimalloc{}
general-purpose allocator~\cite{Leijen2019MimallocFL}. In addition, the
\libname{TLSF}-based \sdrad{} allocator places allocator metadata on the domain
heap, where it could be corrupted by memory errors that occur within sandboxed
code. The \mimalloc{} allocator, similarly, places allocator metadata in a
``backheap" area, that by default would be placed on the domain heap. To harden
the \newsdrad{} allocator, we modified \mimalloc{} to ensure that all
metadata is allocated exclusively from the monitor domain, rather than from a
domain's backheap. This guarantees that allocator metadata is restricted to the
monitor domain, preventing any modifications from sandboxed code. In
\newsdrad{}, the allocator algorithm that is run when \code{malloc} or
\code{sdrad_malloc} is called (\dCSix in \Cref{fig:runtime-arch}), always runs
in the monitor domain (\dCFour).

To harden PKU-based isolation, we extend \sdrad{} with the ERIM binary scanning
method to filter out \code{xrstor} and \code{wrpkru} instructions. In the
following section, we explain our approach to hardening using the syscall
filtering method.

In total, our changes to \sdrad{} amount to around $\approx$1.2K lines of C/C++
code, including
modifications to \mimalloc{}.

\subsection{In-process System Call Filtering.}\label{sec:syscall-filter}

\toolname{} needs a solution for system call interposition
\begin{inparaenum}[1)]
\item to harden PKU-based in-process isolation and
\item to reduce the attack surface for each \toolname{} domain.
\end{inparaenum}
Early research on hardening PKU-based isolation using system-call interposition
can either suffer from high overhead or offer kernel changes or kernel
modules~\cite{Vahldiek-Oberwagner19, Voulimeneas22, Schrammel20}. To address
these limitations, we revisited existing system call interposition methods and
employed a newer approach, \libname{zpoline}~\cite{yasukata23}: a technique for
the x86-64 architecture to intercept system calls. It replaces the two-byte
\code{syscall} or \code{sysenter} instructions with a \code{call rax}
instruction of equal length. Since the system call number is stored in the
\code{rax} register as per the x86-64 call conventions, the \code{call rax}
causes a control-flow transfer to a memory address corresponding to the system
call number, within the first 500 bytes of memory. \libname{zpoline} populates
this range with a series of no-operation instructions, effectively constructing
a \emph{NOP sled}, which eventually leads to a trampoline transferring control
to a user-defined code or a syscall interposer's code. This design ensures that
the rewriting of system call instructions cannot fail, providing a reliable
method for intercepting and customizing system call behavior in applications.
\libname{zpoline} handles static binaries before \code{main}, but it doesn't
support dynamically generated code. To address this, we  hook the \code{mprotect} call that marks memory as executable and scan the memory region for syscall instructions and rewrite them. This allows
\libname{zpoline} to trace syscalls in dynamically generated code as well. Our system call policy enforcement, implemented on top of \libname{zpoline}, consists of approximately 300 lines of code C/C++ code in total.

\paragraph{PKU-Hardening.} \newsdrad{} enforces similar system call filter
policies as previous work on PKU-Based isolation
primitives~\cite{Vahldiek-Oberwagner19, Voulimeneas22}. Using the system call
interpositioning primitive, \libname{zpoline}, (\dCEight in
\Cref{fig:runtime-arch}) we trap system calls from any sandbox domains to a
system call filtering engine in the monitor domain (\dCSeven). The policy engine
applies a default filtering policy that rejects any system calls that attempt to
circumvent the \sdrad{}-enforced PKU policy, such as \code{pkey_mprotect} calls
that attempt to change the PKU keys associated with memory pages, or \code{open}
family systems calls to \code{/proc/self/mem} which would give a sandbox raw
access to the memory of the process. \newsdrad{} also enforces that pages being
dynamically mapped as executable do not contain \code{xrstor} and \code{wrpkru}
instructions which could be used to bypass PKU policies or alter the monitor
state. To ensure this, the monitor scans such pages for the presence of these
instructions, and enforces a Write-XOR-Execute policy on executable pages after
inspection. Also, any system call that attempts to modify the \libname{zpoline}
NOP sled area is rejected by our enforcement mechanism. Systems calls from the
monitor domain itself are exempt from this policy to enable to \sdrad{} domain
management logic (\dCFive) to manage the sandbox PKU policies properly.

\subsection{TCB Size}
At runtime, the trusted computing base includes the memory allocator, the
\sdrad{} monitor, and the generated context-switch code. The memory allocator
contains $\approx$10k lines of code. The \sdrad{} monitor contains $\approx$5k
lines of code. The generated context-switch code contains around 20 to 100 lines
of code for each sandboxed function.


\section{\new{Evaluation}}
\label{sec:eval}

We evaluate \toolname{} along three dimensions: security, usability, and
performance. For security, we demonstrate that \toolname{} mitigates real-world
vulnerabilities. For usability, we apply \toolname{} to real Rust applications
and show that isolation policies can be expressed with little
specification effort. For performance, the dominant cost is sandbox context
switching; so overhead depends on program behavior and boundary placement:
frequent crossings amplify overhead, while boundaries on infrequent paths
(e.g., one-shot parsing) incur little overhead. To understand how overhead
varies across usage patterns, we first measure per-call switching cost in
micro-benchmarks against prior tools and the baseline. With that cost in hand,
we apply \toolname{} to diverse real-world applications with different call
patterns to quantify end-to-end overhead. Specifically, we answer the following
research questions:
\begin{itemize}[leftmargin=*]
\item \textbf{RQ1 (User Effort):} How much effort is required to write specifications for \toolname{}?
\item \textbf{RQ2 (Security):} Can \toolname{} effectively mitigate memory safety vulnerabilities?
\item \textbf{RQ3 (Micro-benchmarks):} How does \toolname{} compare with existing sandboxing tools in controlled benchmarks?
\item \textbf{RQ4 (Real-world Overhead):} What is the runtime overhead when applying \toolname{} to real-world applications?
\end{itemize}

\paragraph{Benchmarks.} Across the evaluation, we use two classes of real-world
Rust programs: crates with known memory-safety advisories and larger
applications/parsing libraries. We randomly selected four Rust crates with
memory safety vulnerabilities recorded in the RustSec Advisory Database:
\libname{transpose}~\cite{transpose}, \libname{rouille}~\cite{rouille},
\libname{elf-rs}~\cite{elfrs}, and \libname{rust-openssl}~\cite{openssl}. We
also applied \toolname{} to larger applications and parsing-heavy libraries,
including \appname{Servo}~\cite{servo},
\appname{ripgrep}~\cite{githubGitHubBurntSushiripgrep}, \appname{oq}~\cite{oq},
and \libname{async-graphql}~\cite{asyncgraphql}.

\paragraph{Experiment setup.} We ran all experiments on a 12-core 13th Gen Intel(R) Core(TM) i7-1360P at 2.10
GHz and 64 GiB of RAM. The machine runs Ubuntu 24.04.3 LTS with version
6.12.25 of the Linux kernel. We disable
hyperthreading on the CPU to reduce measurement noise.

\subsection{RQ1: User Effort for Specification}
\label{sec:eval-effort}

\begin{table}[t!]
\centering
\begin{tabular}{l c c c}
  \toprule
  \textbf{Application} & \textbf{Spec Lines} & \textbf{Sandboxed LOC} & \textbf{Boundary} \\
  \midrule
  \libname{transpose}     & 2 & 671 & 6 \\
  \libname{elf-rs}        & 2 & 1,342 & 1 \\
  \libname{rust-openssl}  & 2 & 49,306 & 28 \\
  \libname{rouille}       & 2 & 10,422 & 1 \\
  \appname{Servo}         & 2 & 134,708 & 1 \\
  \appname{ripgrep}       & 2 & 2,736 & 8 \\
  \appname{oq}            & 2 & 1,767 & 5 \\
  \libname{async-graphql} & 2 & 2,465 & 1 \\
  \bottomrule
\end{tabular}
\caption{Specification size, sandboxed LOC, and public API numbers for benchmark applications or libraries. Notice that here we only count the lines of code for the target crate itself, excluding its dependencies. Although at runtime, the sandboxed code may include code from dependencies as well.}
\label{tab:spec}
\end{table}

We apply the specification strategies described in \Cref{sec:design} to our
benchmark applications. \Cref{tab:spec} summarizes the results.

\paragraph{Library-based isolation.} We apply this strategy to three crates with
known memory-safety vulnerabilities:
\begin{itemize}[leftmargin=*]
\item \libname{transpose} is a matrix library with an
  out-of-bounds write (RUSTSEC-2023-0080~\cite{rustsecRUSTSEC20230080Transpose}) triggered by crafted input dimensions.
\item \libname{elf-rs} is an ELF file parser with a heap buffer overflow
  (RUSTSEC-2022-0079~\cite{RUSTSEC-2022-0079}) when parsing malformed binaries.
\item \libname{rust-openssl} wraps OpenSSL, which has had numerous
  vulnerabilities including CVE-2023-0216~\cite{CVE-2023-0216} (invalid pointer dereference) and
  RUSTSEC-2023-0044~\cite{rustsecRUSTSEC20230044Openssl} (use-after-free).
\end{itemize}

\paragraph{User-based isolation.} We apply this strategy to two multi-tenant
applications:
\begin{itemize}[leftmargin=*]
\item \libname{rouille} is an HTTP server framework. It uses \libname{zlib}
  for compression, which had vulnerabilities including CVE-2018-25032~\cite{CVE-2018-25032},
  CVE-2022-37434~\cite{CVE-2022-37434}, and CVE-2002-0059~\cite{CVE-2002-0059}. We isolate each call to the compression
  function for each user, preventing a compromised request from accessing other
  users' data. Identifying this boundary required tracing through 3 functions
  from the HTTP request handler.
\item \appname{Servo} is Mozilla's experimental browser engine. We isolate the
  JavaScript execution engine for each tab, preventing a malicious script in one
  tab from accessing data in other tabs. We identified the script execution
  component and its entry function from the project's architecture documentation.
\end{itemize}
While this strategy requires understanding the application's data flow, the
effort remains manageable. \gtan{Is there a way of quantifying the amount of effort? Like the number of hours you spent for each?}

\paragraph{Input-oriented isolation.} We apply this strategy to three
applications that process external input:
\begin{itemize}[leftmargin=*]
\item \appname{ripgrep} is a fast grep tool. We isolate its \code{GlobSet}
  struct, which matches file names against user-provided glob patterns. We
  identified this by examining the dependency tree, finding the \libname{globset}
  crate, and locating its public interface struct.
\item \appname{oq} is a command-line tool for querying and transforming data
  in various formats (JSON, YAML, TOML). We isolate the entire \libname{oq}
  crate, which parses untrusted input files.
\item \libname{async-graphql} is a GraphQL server library. We isolate its
  parsing function, \code{parse\_query}, which parses untrusted client
  queries. The identification was straightforward due to the clear function name.
\end{itemize}
These components directly handle external input and benefit from isolation
regardless of whether there are currently known vulnerabilities.

\paragraph{Specification effort.} As shown in \Cref{tab:spec}, each application
requires only \textbf{two lines of specification}. Once the isolation boundary
is identified---whether a crate, struct, or function---specifying it in
\toolname{} is straightforward. The main effort lies in understanding the
application's structure to identify appropriate boundaries, not in writing the
specification itself.

\subsection{RQ2: Security Evaluation}
\label{sec:eval-security}

We evaluate \toolname{}'s security effectiveness through two complementary
approaches: (1) analyzing how \toolname{} would contain known CVEs in real-world
crates, and (2) validating isolation guarantees using proof-of-concept (PoC)
exploits derived from cross-language attack patterns.

\paragraph{Analysis of known vulnerabilities.}
Our benchmark crates contain known vulnerabilities including RUSTSEC-2023-0080~\cite{rustsecRUSTSEC20230080Transpose},
CVE-2018-25032~\cite{CVE-2018-25032}, RUSTSEC-2023-0044~\cite{rustsecRUSTSEC20230044Openssl}, and RUSTSEC-2022-0079~\cite{RUSTSEC-2022-0079} (out-of-bound writes),
as well as RUSTSEC-2023-0044~\cite{rustsecRUSTSEC20230044Openssl}, CVE-2002-0059~\cite{CVE-2002-0059} and CVE-2023-0216~\cite{CVE-2023-0216} (double-free and invalid pointer
dereference). By sandboxing the vulnerable function or crate with \toolname{},
these vulnerabilities would be contained: if triggered, exploits can only access
memory inside the sandbox or will be caught by the low-level isolation
mechanism of \toolname{}, which throws a runtime error when memory outside the sandbox is
accessed.


\paragraph{PoC-based validation.}
To concretely validate \toolname{}'s isolation guarantees, we developed
three proof-of-concept exploits modeling attack patterns from prior
work~\cite{claPaper, mixedBinariesPaper}, including both pure-Rust and
cross-language attacks:

\begin{itemize}[leftmargin=*]
\item \textbf{Lifetime soundness exploit:} A pure-Rust attack exploiting the
  lifetime expansion soundness hole (Rust issue \#25860~\cite{rust25860}) to forge a mutable
  reference to an arbitrary address. This models supply-chain attacks where safe
  Rust can corrupt memory due to compiler unsoundness. When run inside
  \toolname{}'s sandbox, cross-domain writes to the monitor cause memory faults handled by \newsdrad{}, leaving
  the target unchanged.

\item \textbf{Lifetime violation via C FFI~\cite{claPaper}:}
  Rust allocates a buffer, hands its pointer to C, and C frees it behind Rust's
  back; Rust then reads the freed memory, creating a use-after-free. Under
  \toolname{}, when the sandbox attempts to free a buffer allocated by the
  monitor domain, the monitor intercepts and blocks the operation, preventing
  the attacker from invalidating host memory.

\item \textbf{Mixed stack corruption~\cite{mixedBinariesPaper}:}
  When Rust and C coexist, Rust stack frames are not protected by SafeStack; a buffer overflow in C can write past a Rust stack buffer and corrupt adjacent control data
  (e.g., a return address/guard), enabling control-flow hijack. After isolation, the untrusted C
  runs inside a sandbox domain; so the overflow corrupts only the sandbox's own stack; the monitor's control data remains intact, containing the attack's impact.
\end{itemize}

Each PoC includes an \texttt{attack} binary demonstrating the exploit without
isolation, and an \texttt{isolated} binary showing mitigation under \toolname{}.
In all cases, cross-domain memory accesses from sandbox code into monitor memory
result in runtime faults, successfully containing the attack.

\subsection{RQ3: Micro-benchmark Comparison}
\label{sec:eval-micro}

\begin{figure}[t]
  \begin{subfigure}[t]{0.48\textwidth}
    \centering
    \includesvg[width=0.7\linewidth]{fig/svg/figure-all-compress.svg}
    \label{fig:perf1abomonation}
  \end{subfigure}
  \begin{subfigure}[t]{0.48\textwidth}
    \centering
  \includesvg[width=0.7\linewidth]{fig/svg/figure-all-decompress.svg}
    \label{fig:perf1serialization}
  \end{subfigure}
  \caption{Execution Time of \code{compress()} and \code{uncompress()} for \toolname{}, SandCrust, SDRaD-FFI with \texttt{SDRaD}-v1, SDRaD-FFI with \newsdrad{}, Baseline, and Baseline with mimalloc allocator}
  \label{fig:perf1}
\end{figure}

We measure per-call overhead using the compression library \libname{snappy} and the image
codec \libname{libpng}, comparing \toolname{} against
\sdradrustffi{}~\cite{gulmezFriendFoeExploring2023} and
Sandcrust~\cite{lamowskiSandcrustAutomaticSandboxing2017}.

\paragraph{Benchmark construction.} Each benchmark program executes exactly one
sandboxed function call per iteration, measuring total execution time (sandbox
overhead plus workload processing). For snappy, we wrap the \code{compress} and
\code{uncompress} functions, invoking each once per iteration while varying the
input buffer size (1B to 2MB, in powers of two). For libpng, we wrap
\code{decode\_png} and invoke it once per iteration on different test images.
By varying input size, we observe how fixed sandbox overhead amortizes as
workload increases. We perform 100 warm-up iterations
followed by 50,000 timed iterations per configuration, measuring wall-clock
time and computing mean and standard
deviation. We reused the public benchmark artifact from
\sdradrustffi{}~\cite{gulmezFriendFoeExploring2023} to ensure a fair comparison.
\Cref{fig:perf1} presents the overall execution times across different data
sizes.

\paragraph{Baseline comparison.} We evaluated both the standard libc allocator
and mimalloc as baselines. While the difference is negligible for small buffer
sizes, mimalloc provides a clear advantage for larger inputs. Since \toolname{}
uses mimalloc by default, this explains why \toolname{} outperforms the libc
baseline for larger inputs.

\paragraph{SDRaD-FFI comparison.} We measured \sdradrustffi{} with its default TLSF
allocator and our \newsdrad{} variant, which enhances compatibility and
security. As expected, \newsdrad{} incurs additional overhead compared to
\sdradrustffi{}, primarily due to the allocator choice and the syscall
interposition mechanism. Although \sdradrustffi{} uses in-process isolation, it
performs worse than \toolname{} because its data transfers rely on
copying across domains, whereas \toolname{} reduces this overhead
through shared heap access.

\paragraph{Sandcrust comparison.} Sandcrust, which uses processes for
isolation, shows worse performance than \toolname{} due to the overhead of
process context switching and inter-process communication.

\paragraph{Summary.} These micro-benchmarks demonstrate that: (1) the
heap-sharing optimization of \toolname{} effectively reduces sandboxing overhead
compared to copy-based approaches (\sdradrustffi{}); and (2) in-process isolation
performs better than process-based isolation (Sandcrust).

\subsection{RQ4: Real-world Application Overhead}
\label{sec:eval-overhead}

\newcommand{\tworow}[1]{\multirow{-2}{*}{#1}}%
\newcommand{\onerow}[1]{\multirow{1}{*}{#1}}%
\begin{table}[t!]
\centering
\resizebox{0.49\textwidth}{!}{%
  \begin{tabular}{l rr rr}
    \toprule
    & \multicolumn{2}{c}{\textbf{\toolname{}-Copy}} & \multicolumn{2}{c}{\textbf{\toolname{}-Share}}    \\
    \cmidrule(r){2-3} \cmidrule(r){4-5}
    & Min      & Max     & Min     & Max     \\ \midrule

    \rowcolor{gray!10} \libname{rouille}    & \new{37.1}\% & \new{89.0\%} & \new{-0.3}\% & \new{3.0}\%                \\
    \rowcolor{white} \libname{elf-rs}      &   12.7\%  & 96.9\% & 5.4\% & 15.8\%                  \\
    \rowcolor{gray!10} \appname{oq}         & \new{5.7}\% & \new{36.0}\%  & \new{8.4}\%  & \new{33.2}\%                  \\
    \rowcolor{white} \libname{async-graphql} & 6.8\% & 15.8\%  & 5.3\%  & 13.3\%                  \\
    \rowcolor{gray!10} \appname{ripgrep}    &   5\%    & 13.4\%      & 2.9\% & 10.3\%                \\
    \rowcolor{white} \libname{rust-openssl} & \new{3.9}\% & \new{17.7}\%  & \new{0.4}\%  & \new{22.2}\%                \\
    \rowcolor{gray!10} \appname{Servo}        & -       & -       & -1.4\% & 9.4\%               \\
    \rowcolor{white} \libname{transpose}    &  3.3\%& 4.4\%  & 3.0\%  & 4.8\%               \\
    \bottomrule
\end{tabular}} \caption{\toolname{} performance overhead compared to the baseline.
Missing data indicates the application requires data sharing and cannot be
accommodated by the copy mode. Max and min are maximal and minimal overheads
under different workloads. For \libname{rouille}, we use the throughput here.
See \Cref{sec:env} for details.}
\label{tab:eval-summary}
\label{tbl:eval-summary}
\end{table}

We evaluate \toolname{}'s overhead on eight real-world applications and
libraries. We compare \toolname{}-Copy (data copied between domains) and
\toolname{}-Share (heap-sharing optimization). \Cref{tbl:eval-summary}
summarizes the performance overhead ranges. Detailed performance data can be
found in the appendix\Cref{sec:env}.

\paragraph{\appname{Servo}.} We run the Dromaeo JavaScript benchmark suite,
which executes thousands of JavaScript operations per test category. Each
operation triggers a sandbox entry/exit as the JS engine runs in isolation. The
high call frequency (hundreds of calls per millisecond) makes this a stress test
for sandbox transition cost. Threads in Servo use channels to communicate, which
involve heap allocations. We modified the channel implementation by adding 300
lines of code to make it domain-aware, so that \toolname{} can change their
allocation sites to shared domains and allow communication between sandboxes.
Copy mode is not applicable because the inter-thread message channel requires
real memory sharing. Overall, the max overhead is 9.4\%.

\paragraph{\libname{rust-openssl}.}
We use an HTTP library in Rust, attohttpc~\cite{attohttpc} to implement seven request-level tasks (get, post, session, assets, multipart, redirect, stream). Together they cover basic GET and POST, a session with multiple requests and JSON parsing, asset-style transfers with HEAD and ETag
  conditional GET, multipart upload of a ~256 KB payload, redirect chains, and a streamed JSON response; all responses are consumed and checksummed to keep work on the hot path. These tasks are executed against public endpoints (e.g., www.google.com and httpbin.org) to exercise realistic HTTP client behavior. 
  These program can be found in our artifacts (\cref{sec:artifact}).
Each
TLS handshake and data transfer invokes multiple sandboxed OpenSSL functions.
\toolname{}-Copy introduces \new{17.7.0}\% max overhead; \toolname{}-Share
reduces this to \new{22.2}\% max.

\paragraph{\libname{transpose}.} We integrate the sandboxed \code{transpose}
function into \libname{RustFFT}, which uses matrix transpositions as part of
FFT computation. We vary matrix sizes from 64$\times$64 to 4096$\times$4096.
Each FFT operation triggers one sandbox call, and data transfer for \toolname{}-Copy scales
quadratically with matrix size. For performance measurement, we run the FFT algorithm benchmark that comes with \libname{RustFFT}. Both \toolname{}-Copy and \toolname{}-Share have 3\% to 5\% overhead because the FFT computation dominates the running time. 

\paragraph{\libname{rouille}.} We run a local \libname{rouille} server with a configurable thread pool and a single handler
  that returns a fixed‑size response body, and the response passes through \libname{rouille}’s content encoding where compression work is done by the sandboxed function, which calls \libname{zlib}.
  We then drive that server with \appname{wrk}. We configure \appname{wrk} with 8 threads and 200 connections. During the experiement, we run \libname{rouille} with 1,2 and 4 threads.  Each request triggers one sandbox
call to the \code{compress} function. We measure both latency and throughput.
\toolname{}-Copy introduces 89\% max overhead for latency and 77.2\% for throughput; \toolname{}-Share reduces this to
14.9\% max and 3\% max.

\paragraph{\libname{elf-rs}.} We build a \appname{readelf}-like tool that parses
ELF binaries and prints its content. We test on different sizes of binaries. Each file triggers one sandbox call
to the parser, with data transfer proportional to file size. We measure the running time for each file.
\toolname{}-Copy introduces 96.9\% max overhead; \toolname{}-Share reduces this to
15.8\% max.

\paragraph{\appname{oq}.} The \libname{oq} benchmark evaluates the \appname{oq} CLI (a jq-like object query tool) by running a fixed filter over JSON inputs of increasing size. For each sample file (100k–500k JSON) that contains an
  items array, we execute \appname{oq} to parse the input and count the array length. The size of data across the boundary for \toolname{}-Copy is proportional to the input size. For smaller files, the parsing time is dominated by the sandbox transition overhead, therefore, both \toolname{}-Copy and \toolname{}-Share show higher overhead (36.0\% and 33.2\% respectively). When the input size increases, the parsing time dominates, and the overhead decreases (5.7\% and 8.4\% respectively).

\paragraph{\libname{async-graphql}.} We setup a GraphQL server using \libname{async-graphql} with a synthetic dataset (including 2000 entities and relations between them). Then we measure the end-to-end performance with four kinds of queries, covering fragments, pagination/edges, conditional directives, and nested relationships. Each
incoming query triggers one sandbox call to \code{parse\_query}. The parsed AST
is returned to the monitor for execution. \toolname{}-Copy introduces 15.8\% max
overhead; \toolname{}-Share reduces this to 13.3\% max.

\paragraph{\appname{ripgrep}.} We use \appname{ripgrep} to search directory trees with varying numbers
of files (64--512). To check
pattern matches, the methods of the isolated struct \code{GlobSet} are called several times per file , while the array of matched patterns is sent across the boundary. We measure the end-to-end running time. \toolname{}-Copy introduces 13.4\% max
overhead; \toolname{}-Share reduces this to 10.3\% max. However, the 
rest of the overhead is not from the context switching, instead, it 
is because of the \libname{zpoline} syscall interception mechanism 
has non-negligible overhead for the \code{open} syscall, which 
\appname{ripgrep} invokes frequently when searching files.

\paragraph{Summary.} Overhead correlates with call frequency and data transfer
volume. High-frequency call patterns (e.g., \libname{rust-openssl}) stress sandbox transition
costs, while large data transfers (e.g., \libname{elf-rs}) stress the copying mechanism.
\toolname{}-Share outperforms \toolname{}-Copy by eliminating
data copying overhead in most cases. In some cases, the data size across isolation boundary is small and the noise dominates, so that \toolname{}-Share performs similar or slightly worse than \toolname{}-Copy.

\section{Related Work}
\label{sec:related}


\paragraph{In-process isolation}. Several studies have investigated the
compartmentalization of applications based on different underlying primitives,
including hardware-assisted in-process isolation~\cite{Rivera16, Koning17,
Hedayati19, Vahldiek-Oberwagner19, Melara19, Sung20, Lefeuvre21, Schrammel20,
Wang20, Voulimeneas22, Kirth22, Jin22, Chen22, gulmez2023dsn}, process-based
isolation~\cite{lamowskiSandcrustAutomaticSandboxing2017}, and software-based
isolation~\cite{Tan17}. Most of these approaches focus on compartmentalizing
applications written in C or C++~\cite{Koning17, Hedayati19,
Vahldiek-Oberwagner19, Melara19, Sung20, Lefeuvre21, Schrammel20, Wang20,
Voulimeneas22, Jin22, Chen22, gulmez2023dsn}.  Additionally, several studies
specifically address isolating unsafe code from safe code for Rust
application\citeUnsafeIso{}. However, those approaches implement conservative
compartmentalization policies that label any objects potentially used by
untrusted code as unsafe. This can result in edge cases where most allocations
are relegated to the unsafe domain, which in turn diminishes the intended
benefits of compartmentalization. Also,
\sdradrustffi{}~\cite{gulmezFriendFoeExploring2023} and
Sandcrust~\cite{lamowskiSandcrustAutomaticSandboxing2017} focus on providing a
finer-grained isolation boundary with manual function annotation, but they do
not scale effectively for larger isolation boundaries, such as an entire Rust
crate. In contrast, \toolname{} provides a highly automated solution for sandboxing
both safe and unsafe code in Rust applications.

\paragraph{Language-based isolation}.
One idea of language-based isolation is to provide language constructs that
allow developers to specify isolation boundaries and manage sandboxes.  For
example, Enclosure~\cite{ghosnEnclosureLanguagebasedRestriction2021} allows
developers to specify capabilities of libraries. GOTEE~\cite{ghosn2019secured}
puts a Go routine into an Intel SGX enclave and automatically extracts sandboxed
code and data from the original code. Both systems allow the sandbox boundary
based on  one existing
language construct (libraries or Go Routines). \toolname{} further pushes this
direction by providing flexible boundaries such as crates, functions and
types. Also, prior work target Go applications, while \toolname{} targets Rust applications.  Another idea is to provide a sandbox interface in the language or through library, to
support sandboxed execution of external code. JavaScript engine
V8~\cite{GoogleV8} can be used as a virtual machine that runs user code, whose
interaction with the outside world is managed by V8.
WebAssembly~\cite{WebAssembly} provides a sandboxed execution environment for
web applications by including bounds checking in the semantics of the
instruction set.  Libsandbox~\cite{githubGitHubOpenJudgeSandbox} provides APIs
in C/C++ and Python for executing and profiling single process programs in a
sandbox.  However, these tools are designed to execute foreign code in a one-off
manner with a high cost of context switching. Furthermore, retrofitting legacy
code into these APIs is not desirable for large-code bases. In contrast, \toolname{} leverages
existing syntax boundaries to alleviate the effort needed for compartmentalization.

\paragraph{Automated privilege separation}.
There is a long line of work that employs program analysis to separate a program
into multiple partitions automatically (e.g.,
Privtrans~\cite{brumley2004privtrans}, ProgramCutter~\cite{wu2013automatically},
SeCage~\cite{liu2015thwarting}, Glamdring~\cite{lind2017glamdring},
PtrSplit~\cite{liu2017ptrsplit},
Jif/split~\cite{Zdancewic2002secure,zheng2003using},
Swift~\cite{chong2007secure}, $\mu$SCOPE~\cite{roessler2021muscope} ). These
systems target C/C++ or Java applications, instead of Rust applications. In
addition, their program analysis is mostly focused on computing partitioning
boundaries.  Our system's static analysis instead computes allocation sites that
can produce cross-boundary objects to reduce overhead when objects are shared
through share-memory regions. There has also been research of reducing the
overhead of copying cross-boundary objects. E.g., KSplit~\cite{huang2022ksplit}
copies only necessary fields of cross-boundary objects, where the set of
necessary fields is computed by static analysis.

\section{Discussion \& Future Work}\label{sec:discussion}


\paragraph{In-process isolation.} Like other in-process isolation relying on Intel \gls{mpk}, \toolname{} is constrained by the hardware limit of 16 distinct protection domains. As \toolname{}'s specification and analysis can work independently of the underlying enforcement mechanism, future work could explore replacing MPK with alternative in-process isolation schemes, such as CHERI-based hardware capabilities~\cite{cheri} or \gls{sfi}~\cite{Tan17}.

\paragraph{System Call Filtering Alternatives}.
As we discuss in Section~\ref{sec:syscall-filter}, \toolname{} integrates the zpoline system call interpositioning as part of SDRaD-\toolname{} to harden PKU-based isolation and to address \cha{4}.
While SDRaD-\toolname{} is fully compatible with conventional system call filtering techniques used in prior PKU-based isolation approaches~\cite{Vahldiek-Oberwagner19, Voulimeneas22}, these techniques typically require kernel modifications to improve their efficiency. Without kernel support, they tend to incur significant performance overhead.

\paragraph{System Call Filtering Enhancements}. Currently, \toolname{}'s isolation
only focuses on memory isolation. However, its system call interpositioning
mechanism could also be used to enforce more system call policies. For example, we
can extend the specification to allow developers to deny certain system call for a
particular sandbox. We leave this as future work.

\section{Conclusion}\label{sec:conclusion}
We presented \toolname{}, a novel compiler-based tool that analyzes and instruments Rust code to provide component isolation; we believe that our open-source release of \toolname{} paves the way for future advancements in securing the Rust ecosystem.

\section*{Ethical Considerations}
This research adheres fully to established ethical guidelines. It does not result in any tangible harm nor does it infringe upon human rights. All code and content presented in this paper were originally developed by the authors and are free from plagiarism or unauthorized use of external sources.

\section*{Generative AI Usage}
We used OpenAI Codex to assist in the debugging process of our Rust codebase and
to generate our evaluation scripts/workloads. The debugging output can be
straightforwardly verified. The evaluation scripts/workloads were reviewed and
modified by the authors to ensure correctness. We also use Google's Gemini to
help searching for related work and benchmarks. The search results were further
examined by the authors to ensure relevance. For grammar and spelling checking,
we used GitHub Copilot.


\bibliographystyle{acm}
\bibliography{reference, local, vul}

\newpage
\appendix







\section{Open Science}\label{sec:artifact}
The artifacts are available at \url{https://anonymous.4open.science/r/SandCell-Artifact-2D4E}.
The artifacts include the implementation of the tool, \toolname{}, and
all the benchmarks used in the evaluation, along with scripts to reproduce the results. The instructions for reproducing the results are included in the
README file in the artifact repository.

\section{Supplementary Measurement}\label{sec:env}
We list all the data from the evaluation below. For readability, we include both the 
plots and tables for these experiments.

\begin{figure*}[t!]
\centering
\begin{subfigure}[b]{0.32\textwidth}
  \centering
  \includesvg[width=\linewidth]{fig/svg/elf-rs.running-time.svg}
  \caption{Running Time of \libname{elf-rs}}
  \label{fig:elfrs}
\end{subfigure}%
\begin{subfigure}[b]{0.32\textwidth}
  \centering
  \includesvg[width=\linewidth]{fig/svg/transpose.running-time.svg}
  \caption{Running Time of \libname{transpose}}
  \label{fig:transpose}
\end{subfigure}%
\begin{subfigure}[b]{0.32\textwidth}
  \centering
  \includesvg[width=\linewidth]{fig/svg/ripgrep.running-time.svg}
  \caption{Running Time of \libname{ripgrep}}
  \label{fig:ripgrep}
\end{subfigure}
\begin{subfigure}[b]{0.32\textwidth}
  \centering
  \includesvg[width=\linewidth]{fig/svg/async-graphql.running-time.svg}
  \caption{Running Time of \libname{async-graphql}}
  \label{fig:async-graphql.running-time.svg}
\end{subfigure}%
\begin{subfigure}[b]{0.32\textwidth}
  \centering
  \includesvg[width=\linewidth]{fig/svg/oq.running-time.svg}
  \caption{Running Time of \libname{oq}}
  \label{fig:async-graphql.running-time.svg}
\end{subfigure}%
\begin{subfigure}[b]{0.32\textwidth}
  \centering
  \includesvg[width=\linewidth]{fig/svg/attohttpc.running-time.svg}
  \caption{Running Time of \libname{attohttpc}}
  \label{fig:async-graphql.running-time.svg}
\end{subfigure}%
\caption{Performance benchmarks.}
\label{fig:allbenchmarks}
\end{figure*}

\begin{table*}[h]
\centering
\begin{tabular}{lcccccc}
  \toprule
  Image & Baseline & \sandcrust{} & \toolname{} & Overhead of  & Overhead of & Overhead  \\
  Size &  &  &  & \toolname{}  & \sdradrustffi{} & \sdradrustffi{}  \\
  \midrule
  5.5K & $160.17$ & +183.06\% & $164.30$ & +2.58\% & $176.28$ & +10.06\% \\
  63K & $2391.45$ & +225.99\% & $2389.73$ & -0.07\% & $2506.35$ & +4.80\% \\
  378K & $5797.16$ & +64.21\% & $5844.68$ & +0.82\% & $5904.48$ & +1.85\% \\
  874K & $8961.68$ & +129.44\% & $8943.76$ & -0.20\% & $9306.31$ & +3.85\% \\
  \bottomrule
\end{tabular}
\caption{The Percentage of Overhead for Sandboxing \libname{libpng} Compared to Baseline.}
\label{tab:libpng}
\end{table*}

\begin{table*}[t]
\centering
\begin{tabular}{lccccc}
  \toprule
  Matrix Size & Baseline & \toolname{}-Share & Overhead of \toolname{}-Share & \toolname{}-Copy & Overhead of \toolname{}-Copy \\
  \midrule
  64 & $512.370(22.420\sigma)$ & $537.210(21.470\sigma)$ & $4.8\%$ & $535.010(22.300\sigma)$ & $4.4\%$ \\
  80 & $957.590(30.700\sigma)$ & $993.410(30.630\sigma)$ & $3.7\%$ & $992.070(50.790\sigma)$ & $3.6\%$ \\
  96 & $1476.160(38.400\sigma)$ & $1520.240(36.120\sigma)$ & $3.0\%$ & $1525.440(54.950\sigma)$ & $3.3\%$ \\
  112 & $1892.640(65.620\sigma)$ & $1961.020(45.780\sigma)$ & $3.6\%$ & $1954.820(64.460\sigma)$ & $3.3\%$ \\
  128 & $2046.130(44.300\sigma)$ & $2125.820(57.530\sigma)$ & $3.9\%$ & $2130.330(45.710\sigma)$ & $4.1\%$ \\
  \bottomrule
\end{tabular}
\caption{\libname{RustFFT} with \libname{transpose}: Running Time (milliseconds) on Square Matrices of Different Lengths.}
\label{tab:tranpose}
\end{table*}

\begin{table*}[t]
\begin{tabular}{lccccc}
\toprule
Task & Baseline & \toolname{}-Share & Overhead of \toolname{}-Share & \toolname{}-Copy & Overhead of \toolname{}-Copy \\
\midrule
get & $51.148(7.324\sigma)$ & $55.097(3.078\sigma)$ & $7.7\%$ & $57.339(6.320\sigma)$ & $12.1\%$ \\
post & $51.061(5.741\sigma)$ & $62.377(21.081\sigma)$ & $22.2\%$ & $60.089(13.794\sigma)$ & $17.7\%$ \\
session & $182.633(21.552\sigma)$ & $195.319(56.078\sigma)$ & $6.9\%$ & $200.248(33.824\sigma)$ & $9.6\%$ \\
multipart & $61.411(4.607\sigma)$ & $74.763(12.469\sigma)$ & $21.7\%$ & $72.036(17.354\sigma)$ & $17.3\%$ \\
redirect & $330.272(72.470\sigma)$ & $331.614(35.337\sigma)$ & $0.4\%$ & $343.209(43.260\sigma)$ & $3.9\%$ \\
stream & $54.910(8.434\sigma)$ & $61.294(14.557\sigma)$ & $11.6\%$ & $62.426(17.467\sigma)$ & $13.7\%$ \\
\bottomrule
\end{tabular}
\caption{HTTP Client backed by \libname{openssl}: Running Time on Different Tasks.} 
\label{tab:tranpose}
\end{table*}

\begin{table*}[ht]
\centering
\begin{tabular}{lccccc}
  \toprule
  File Count (8MB each) & Baseline & \toolname{}-Share & Overhead of \toolname{}-Share & \toolname{}-Copy & Overhead of \toolname{}-Copy \\
  \midrule
  64 & $119.476(11.456\sigma)$ & $131.738(13.340\sigma)$ & $10.3\%$ & $135.476(9.656\sigma)$ & $13.4\%$ \\
  128 & $223.320(12.575\sigma)$ & $229.760(3.558\sigma)$ & $2.9\%$ & $233.460(1.739\sigma)$ & $4.5\%$ \\
  256 & $424.456(23.881\sigma)$ & $447.404(33.566\sigma)$ & $5.4\%$ & $453.131(28.304\sigma)$ & $6.8\%$ \\
  512 & $833.026(35.350\sigma)$ & $857.149(40.615\sigma)$ & $2.9\%$ & $874.843(42.189\sigma)$ & $5.0\%$ \\
  \bottomrule
\end{tabular}
\caption{\appname{ripgrep}'s Running Time  on Different Number of Files. }
\end{table*}

\begin{table*}[t]
\centering
\begin{tabular}{lccccc}
  \toprule
  File Size & Baseline & \toolname{}-Share & Overhead of \toolname{}-Share & \toolname{}-Copy & Overhead of \toolname{}-Copy \\
  \midrule
  47 & $9.532(1.680\sigma)$ & $10.601(1.442\sigma)$ & $11.2\%$ & $10.928(1.347\sigma)$ & $14.6\%$ \\
  195 & $11.039(1.704\sigma)$ & $11.631(1.576\sigma)$ & $5.4\%$ & $12.437(1.710\sigma)$ & $12.7\%$ \\
  1192 & $22.910(4.491\sigma)$ & $24.441(3.216\sigma)$ & $6.7\%$ & $33.654(2.131\sigma)$ & $46.9\%$ \\
  5805 & $61.925(2.556\sigma)$ & $71.721(2.248\sigma)$ & $15.8\%$ & $121.957(2.723\sigma)$ & $96.9\%$ \\
  \bottomrule
\end{tabular}

\caption{Parsing ELF Files with \libname{elf-rs} Running Time}
\label{table:elfperf}
\end{table*}

\begin{table*}[htbp]
\begin{tabular}{lcccccc}
\toprule
\multirow{2}{*}{Pool Size} & \multicolumn{2}{c}{Baseline} & \multicolumn{2}{c}{\toolname{}-Share} & \multicolumn{2}{c}{\toolname{}-Copy} \\ & Latency (us) & Requests/sec & Latency (us) & Requests/sec & Latency (us) & Requests/sec \\
\midrule
1 & 8820.0 & 1584.41 & 10130.0 & 1577.41 & 13830.0 & 1155.43 \\
2 & 5030.0 & 3180.91 & 4380.0 & 3191.21 & 7360.0 & 2171.07 \\
4 & 2540.0 & 6302.12 & 2610.0 & 6115.68 & 4500.0 & 3335.04 \\
\bottomrule
\end{tabular}
\caption{\libname{rouille} Performance across Different Thread Pool Sizes.}
\end{table*}

\begin{table*}[h]
\begin{tabular}{lccccc}
\toprule
Query & Baseline & \toolname{}-Copy & Overhead of \toolname{}-Copy & \toolname{}-Share & Overhead of \toolname{}-Share \\
\midrule
static.graphql & $34.173(3.280\sigma)$ & $39.568(5.655\sigma)$ & $15.8\%$ & $38.572(2.971\sigma)$ & $12.9\%$ \\
dashboard.graphql & $29.245(0.882\sigma)$ & $33.661(3.156\sigma)$ & $15.1\%$ & $33.132(0.878\sigma)$ & $13.3\%$ \\
batch.graphql & $30.781(2.883\sigma)$ & $34.585(1.793\sigma)$ & $12.4\%$ & $34.121(0.750\sigma)$ & $10.8\%$ \\
connections.graphql & $72.355(2.465\sigma)$ & $77.309(5.699\sigma)$ & $6.8\%$ & $76.200(2.146\sigma)$ & $5.3\%$ \\
\bottomrule
\end{tabular}
\caption{\libname{async-graphql} Performance Benchmark across Different Query Workloads.}
\end{table*}

\begin{table*}[h]
\begin{tabular}{lccccc}
\toprule
Input & Baseline & \toolname{}-Share & Overhead of \toolname{}-Share & \toolname{}-Copy & Overhead of \toolname{}-Copy \\
\midrule
100k & $13.143(2.519\sigma)$ & $17.511(2.717\sigma)$ & $33.2\%$ & $17.871(3.466\sigma)$ & $36.0\%$ \\
200k & $22.894(3.975\sigma)$ & $24.826(2.458\sigma)$ & $8.4\%$ & $24.206(1.069\sigma)$ & $5.7\%$ \\
336k & $33.901(5.077\sigma)$ & $37.675(3.065\sigma)$ & $11.1\%$ & $37.424(3.467\sigma)$ & $10.4\%$ \\
400k & $37.334(2.568\sigma)$ & $41.679(2.327\sigma)$ & $11.6\%$ & $40.845(3.113\sigma)$ & $9.4\%$ \\
500k & $45.952(7.905\sigma)$ & $50.595(3.499\sigma)$ & $10.1\%$ & $50.986(3.096\sigma)$ & $11.0\%$ \\
\bottomrule
\end{tabular}
\caption{\libname{oq} Performance Benchmark for Different Input Size.}
\end{table*}

\begin{table*}[h]
\begin{tabular}{lc}
\toprule
Benchmark & Overhead of \toolname{}-Share \\
\midrule
Sunspider & -0.007\% \\ 
v8 & 1.650\% \\
jzlib & -1.432\% \\
dom & -0.131\% \\
dromaeo & 9.360\% \\
\bottomrule
\end{tabular}
\caption{\appname{Servo} Average Overhead on each Testsuite of the Dromaeo Benchmark.}
\end{table*}

\end{document}